\documentclass[%
reprint,
superscriptaddress,
amsmath,amssymb,aps, physrev,
]{revtex4-2}

\usepackage{color}
\usepackage{xcolor}
\usepackage{graphicx}
\usepackage{dcolumn}
\usepackage{siunitx}
\usepackage{bm}
\usepackage[version=4]{mhchem}
\usepackage{hyperref}
\hypersetup{colorlinks=true,linkcolor=blue,citecolor=blue,urlcolor=blue}

\usepackage[colorinlistoftodos]{todonotes}

\usepackage{soul} 

\begin{document}
\title[Orbital-selective band evolution and out-of-plane correlation in the FeGe-family kagome antiferromagnet $\text{ScFe}_6\text{Ge}_6$]{Orbital-selective band evolution and out-of-plane correlation in the FeGe-family kagome antiferromagnet $\text{ScFe}_6\text{Ge}_6$}

\author{Jae Hyuck Lee}
\thanks{These authors contributed equally.}
\affiliation{Department of Physics and Astronomy, Seoul National University, Seoul 08826, Republic of Korea}

\author{Ze Yan}
\thanks{These authors contributed equally.}
\affiliation{State Key Laboratory of Quantum Functional Materials, School of Physical Science and Technology, ShanghaiTech University, Shanghai 201210, China}

\author{Tongrui Li}
\affiliation{National Synchrotron Radiation Laboratory, University of Science and Technology of China, Hefei 230026, China}

\author{Yichen Yang}
\affiliation{National Key Laboratory of Materials for Integrated Circuits, Shanghai Institute of Microsystem and Information Technology (SIMIT), Chinese Academy of Sciences, Shanghai 200050,
China}

\author{Dirk Wulferding}
\affiliation{Department of Physics and Astronomy, Sejong University, Seoul 05006, Republic of Korea}

\author{Jongkeun Jung}
\affiliation{Hydrogen·Fuel Cell Research Center, Korea Institute of Science and Technology (KIST), Seoul 02792, Republic of Korea}
\affiliation{Synchrotron Radiation Research Center, National Institutes for Quantum Science and Technology, Sayo, 679-5148, Japan}

\author{Zhicheng Jiang}
\affiliation{National Synchrotron Radiation Laboratory, University of Science and Technology of China, Hefei 230026, China}

\author{Mao Ye}
\affiliation{Shanghai Synchrotron Radiation Facility, Shanghai Advanced Research Institute, Chinese Academy of Sciences, Shanghai 201210, China}

\author{Zhengtai Liu}
\affiliation{Shanghai Synchrotron Radiation Facility, Shanghai Advanced Research Institute, Chinese Academy of Sciences, Shanghai 201210, China}

\author{Changyoung Kim}
\email{changyoung@snu.ac.kr}
\affiliation{Department of Physics and Astronomy, Seoul National University, Seoul 08826, Republic of Korea}

\author{Soohyun Cho}
\email{scho@scnu.ac.kr}
\affiliation{Department of Physics Education, Sunchon National University, Sunchon 57922, Republic of Korea}

\author{Yanfeng Guo}
\email{guoyf@shanghaitech.edu.cn}
\affiliation{State Key Laboratory of Quantum Functional Materials, School of Physical Science and Technology, ShanghaiTech University, Shanghai 201210, China}
\affiliation{ShanghaiTech Laboratory for Topological Physics, ShanghaiTech University, Shanghai 201210, China}

\author{Dawei Shen}
\email{dwshen@ustc.edu.cn}
\affiliation{National Synchrotron Radiation Laboratory and School of Nuclear Science and Technology, University of Science and Technology of China, Hefei 230026, China}

\date{\today}

\begin{abstract}
In strongly correlated materials such as high-temperature superconductors, the relation between charge density wave (CDW) order and magnetism remains an important unresolved problem. FeGe is the first kagome metal known to exhibit CDW order deep within an antiferromagnetic state, accompanied by an unconventional evolution of lattice symmetry. To elucidate the general conditions governing such spin-correlated CDW order, we investigate ScFe$_6$Ge$_6$, which lacks CDW order and therefore exhibits reduced involvement of charge degrees of freedom while retaining other properties of FeGe. Instead of a CDW order, ScFe$_6$Ge$_6$ undergoes a magnetic transition at $T^*$ = 195 K. Across this transition, angle-resolved photoemission and Raman spectroscopy reveal orbital-selective behavior confined to a single kagome Dirac band, together with electron–phonon and magnetoelastic coupling to an out-of-plane phonon mode. These results suggest that orbital-selective physics and out-of-plane correlations play enhanced roles in realizing spin-correlated CDW order in magnetic kagome metals.

\vspace{0.5em}
\end{abstract}

\maketitle

\begin{center}
\noindent\textbf{I. INTRODUCTION}
\end{center}

Kagome materials have attracted enormous attention over the past decade as a platform for studying the complex interplay among spin, charge, orbital, and lattice degrees of freedom, as well as the coexistence of topology and strong correlations~\cite{kagomeRev0,kagomeRev1,kagomeRev2,kagomeRev3,massivedirac1,massivedirac2}. Based on the unique corner-sharing triangular frustrated lattice, kagome systems host three distinct electronic band features: a singular flat band (FB), Dirac crossing (DC) of bands at the K point, and a Van Hove Singularity (VHS) at the M point~\cite{kagomeRev0,kagomeRev1,kagomeRev2,kagomeRev3,SFBRev,Flatband,massivedirac1,massivedirac2,VHS1,VHS2,VHS3,FG2MingNatphy,FG1Nat}. Depending on which feature of the band structure is located close to the Fermi level ($E_F$), kagome lattices have been conjectured, and later reported to host a wide variety of diverse quantum states ~\cite{kagomeRev0,kagomeRev1,kagomeRev2,kagomeRev3,SFBRev,Flatband,massivedirac1,massivedirac2,VHS1,VHS2,VHS3,FG1Nat,FG2MingNatphy}.
In particular, when VHSs lie near $E_F$, they induce electronic instabilities that can lead to the formation of charge density waves (CDW), nematic phases, and superconductivity, as observed in the celebrated $A$V$_3$Sb$_5$ ("135") (A = K, Rb, Cs) compounds~\cite{kagomeRev1,kagomeRev2,kagomeRev3,VHS1,VHS2,VHS3}. 

Besides this prototypical "135" family of kagome metals, FeGe realizes a rare coexistence of magnetism and CDW while displaying VHSs at the $E_F$~\cite{FG2MingNatphy,FG1Nat}. A short-range 2$\times$2$\times$2 CDW order was reported deep \textit{within} the A-type antiferromagnetic (AFM) phase, contrary to other strongly correlated materials such as cuprates and nickelates~\cite{Tc1,Nickelate,NickelREv}, pointing to an intertwining of spin and charge order~\cite{FG0,FG2MingNatphy,FG1Nat,FG5-1pengnatcomm,FG7prl,FG9vacancypeng,FG4SciadMing,FG8donglai,FG12hundmetal,FG10Raman,FG5PengarX,FG8-1donglai,buildingblock,FG11spinphononcoupling,FG13IRDressel,FG3BernavNatcom,FG6applphys}. In addition, a disputed secondary magnetic phase transition occurs at a lower temperature, which displays a strong correlation with the CDW phase~\cite{FG9vacancypeng}. It shows a purported incommensurate spin order that was initially assigned to spin-canting behavior away from the c-axis or, more recently, a spin density wave (SDW)~\cite{FG4SciadMing,FG5PengarX,FG9vacancypeng,FG5-1pengnatcomm,FG7prl,FG0}. 
Despite substantial efforts to uncover a generic mechanism for CDW formation in a magnetic state using FeGe as a model system, the microscopic origin of the incommensurate spin phase and CDW remains unclear~\cite{Competingphase}. Nevertheless, a growing consensus attributes CDW order to a phonon-instability-driven out-of-plane dimerization of the Ge1-site atoms, which lie within the Fe kagome plane~\cite{FG5PengarX,FG8-1donglai,FG10Raman,FG12hundmetal}. Accordingly, considerable effort has been devoted to elucidating the origin of the CDW in FeGe by controlling the Ge-vacancy concentration via annealing, thereby tuning the Ge dimerization~\cite{FG8donglai,FG4SciadMing,FG9vacancypeng}.

\begin{figure*}[htpb]
    \centerline{\includegraphics[width=15cm]{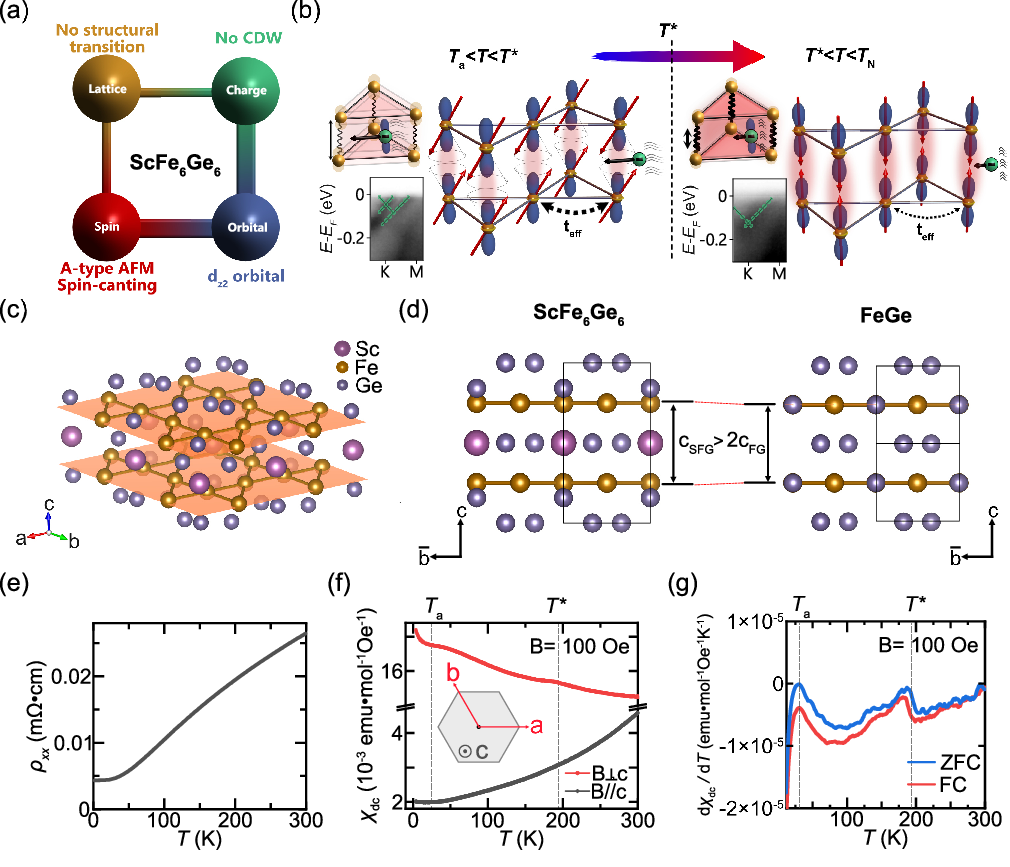}}
    \caption{\textbf{Basic characterization of ScFe$_6$Ge$_6$.} \textbf{(a)} Summary of the ground state of ScFe$_6$Ge$_6$ for each degrees of freedom. \textbf{(b)} Schematic depiction of the interaction between out-of-plane lattice vibrations, magnetic texture, and out-of-plane orbital electronic states in ScFe$_6$Ge$_6$. The magnetic order is represented by the red arrows and the induced band structure evolution is given by the green dotted lines. \textbf{(c)} Lattice structure of the ScFe$_6$Ge$_6$, where the orange planes highlight the kagome layers. \textbf{(d)} Side view of ScFe$_6$Ge$_6$ and FeGe. The rectangular black line marks the unit cell of each compound. The distances between the Fe kagome planes are compared between ScFe$_6$Ge$_6$ and FeGe. \textbf{(e)} Temperature dependent longitudinal resistivity. \textbf{(f)} Field cooled (FC) in-plane (red) and out-of-plane (black) temperature dependent magnetic susceptibility data. \textbf{(g)} First derivative of the in-plane magnetic susceptibility data. The light blue and red lines each represent the zero field cooled (ZFC) and FC data, respectively. The vertical dotted lines indicate the transition temperatures near $T_a$ $\approx$ 30\,K and $T^*$ $\approx$ 195\,K.}
    \label{fig1}
\end{figure*}

In this work, we adopt a different approach to uncover the mechanism underlying spin-correlated CDW order by investigating the FeGe-family derivative compound ScFe$_6$Ge$_6$, thereby extending the general framework established for FeGe-related kagome systems~\cite{buildingblock}. While ScFe$_6$Ge$_6$ displays the same A-type AFM order below $T_N$, we report the emergence of a new magnetic phase transition at $T^*$ in ScFe$_6$Ge$_6$, effectively replacing the CDW transition of FeGe. In contrast to FeGe, where charge ordering drives prominent phase features, the transition in ScFe$_6$Ge$_6$ is purely magnetic, with no accompanying structural distortion or CDW transition as summarized in Fig.~\ref{fig1}(a). Notably, we identify the magnetic phase transition possibly linked to a canted antiferromagnetic ordering, which induces significant band separation. Indeed an orbital-selective band shift across $T^*$ is revealed by angle-resolved photoemission spectroscopy (ARPES). The lattice degree of freedom is further investigated to uncover out-of-plane lattice vibration coupled to the magnetic transition and electronic band structure (Fig.~\ref{fig1}(b)). These results underscore the importance of orbital-selectivity and out-of-plane fluctuations in determining the presence or absence of novel CDW in FeGe-related kagome magnets. The interplay between spin, electronic, and lattice degrees of freedom provides a common underlying mechanism responsible for the rich physics in FeGe-related compounds. 

\begin{figure*}
    \includegraphics[width=16cm]{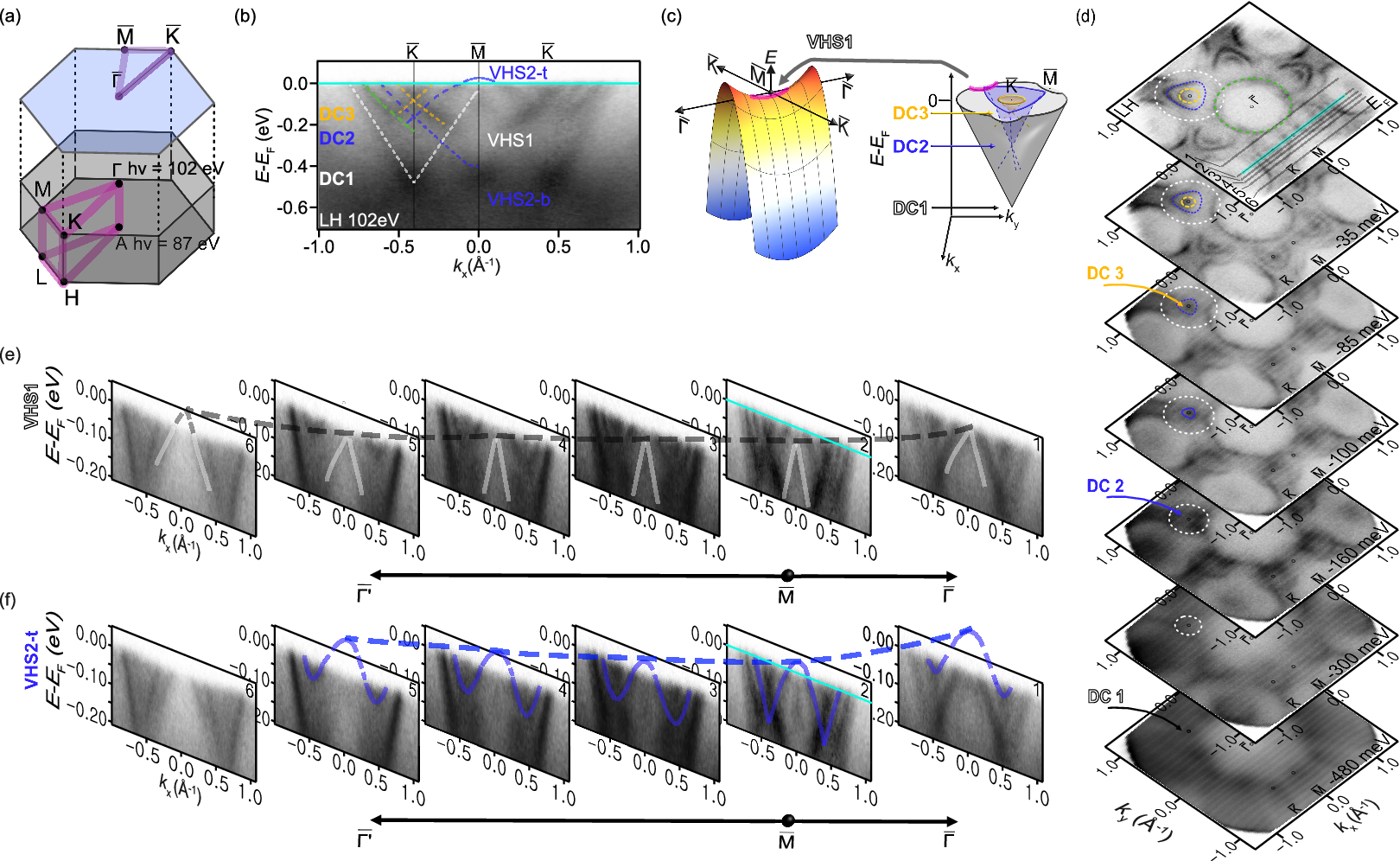}
    \centering
    \caption{\textbf{Electronic band structure of ScFe$_6$Ge$_6$ at 9K.} \textbf{(a)} 3D and projected 2D Brillouin zone (BZ) with each photon energy corresponding to a high-symmetry point. All ARPES data in this paper were taken with a photon energy of 102 eV. \textbf{(b)} Spectral image measured along the $\bar{K}-\bar{M}-\bar{K}$ direction under LH polarization. The white, blue, and orange dotted lines are guides to the eye for the three sets of kagome Dirac bands near $E_F$, while the green dotted line is the hole band (HB) centered at the $\bar{\Gamma}$. Each of the Dirac crossings and Van Hove Singularities (VHSs) is marked accordingly. \textbf{(c)} Schematic diagram of the VHSs and Dirac cones of the kagome band structure in ScFe$_6$Ge$_6$. \textbf{(d)} Constant energy contour plots measured with LH polarization. White, blue, and orange dotted lines centered around $\bar{K}$ indicate the three kagome Dirac cones, and each Dirac crossing point is highlighted with arrows. The green dotted circle is the HB. \textbf{(e),(f)} Stacks of cuts along constant $k_y$ values, perpendicular to the $\bar{\Gamma}-\bar{M}-\bar{\Gamma}$ direction, as indicated by the six black lines on the Fermi surface of (d). The black and blue thick dashed lines in (e) and (f) each highlight VHS1 and VHS-t, respectively. The $\bar{K}-\bar{M}-\bar{K}$ cut is all marked with cyan lines.}
    \label{fig2}
\end{figure*}

\begin{center}
\noindent\textbf{II. RESULTS}
\end{center}
 
A detailed characterization of the material properties was carried out on single crystals of ScFe$_6$Ge$_6$, and our results are consistent with previously reported data from polycrystalline samples~\cite{RFGref1,RFGref2}. Single-crystal X-ray diffraction (SXRD) confirmed that ScFe$_6$Ge$_6$ crystallizes in the HfFe$_6$Ge$_6$-type structure with P6/mmm symmetry (Fig.~1(c)), which is also evident from the naturally formed hexagonal facets of the single crystals (Fig. S1)~\cite{RFGref1,RFGref2}. The unit cell of ScFe$_6$Ge$_6$ comprises two Fe kagome planes forming covalent bonds with the A-site Sc triangular lattice situated between them~\cite{buildingblock}. Compared to the parent compound FeGe (8.10~\AA), the distance between the kagome planes is larger for ScFe$_6$Ge$_6$ (8.12~\AA; Fig.~\ref{fig1}(d)), while the incorporation of Sc approximately doubles the c-axis lattice constant. Notably, unlike in FeGe, where the Ge atoms lie within the Fe kagome plane, the Ge atoms in ScFe$_6$Ge$_6$ are displaced away from the kagome plane.

Figure 1(e) shows the temperature dependent longitudinal resistivity data in zero magnetic field over a temperature range of 5$-$300 K. A resistivity of a few tens of $\mu\Omega \cdot$cm was measured, indicative of good metallic behavior. The resistivity curve also follows $n\,=\,5$ behavior at low temperatures, consistent with a conventional metal in which electron-phonon scattering dominates until the resistivity saturation regime sets in at high temperatures. No apparent phase transition is observed in the resistivity curve, indicating the absence of any long-range charge order, contrary to FeGe. 

On the other hand, the temperature-dependent magnetic susceptibility $\chi(T)$ of ScFe$_6$Ge$_6$ exhibits features of multiple magnetic phase transitions as shown in Fig.~\ref{fig1}(f). The magnetic field was set to 100 Oe perpendicular and parallel to the c-axis (Fig.~\ref{fig1}(f)). ScFe$_6$Ge$_6$ and other $R$Fe$_6$Ge$_6$ compounds show A-type AFM ordering below $T_N$ similar to the parent compound FeGe~\cite{RFGref1,RFGref2}. Indeed, strong magnetic anisotropy was observed between the in-plane and out-of-plane susceptibility in Fig.~\ref{fig1}(f). Such large anisotropy is a common feature of A-type antiferromagnetic 2D kagome systems~\cite{RFGref1,AnisotropyM}. Moreover, the in-plane susceptibility is one order of magnitude larger than the out-of-plane results at low temperatures, indicating the c-axis to be the easy axis, as expected from a typical A-type AFM compound. 

Aside from the $T_N$ (487 K)~\cite{RFGref1} which is above the temperature range dealt within this study, two additional transitions are identified at  $T_a = 30$ K and $T^* = 195$ K in the in-plane susceptibility result (Fig.~\ref{fig1}(g)). At $T_a$, there is a sudden upturn in the magnetic susceptibility, which deviates from the behavior of a simple uniaxial AFM where the in-plane susceptibility should remain constant at low temperatures. Previous studies on powder samples have also reported this anomaly, and it appears to be a common trait in all 166-FeGe-family compounds~\cite{RFGref1}. 

 \begin{figure*}
    \includegraphics[width=17cm]{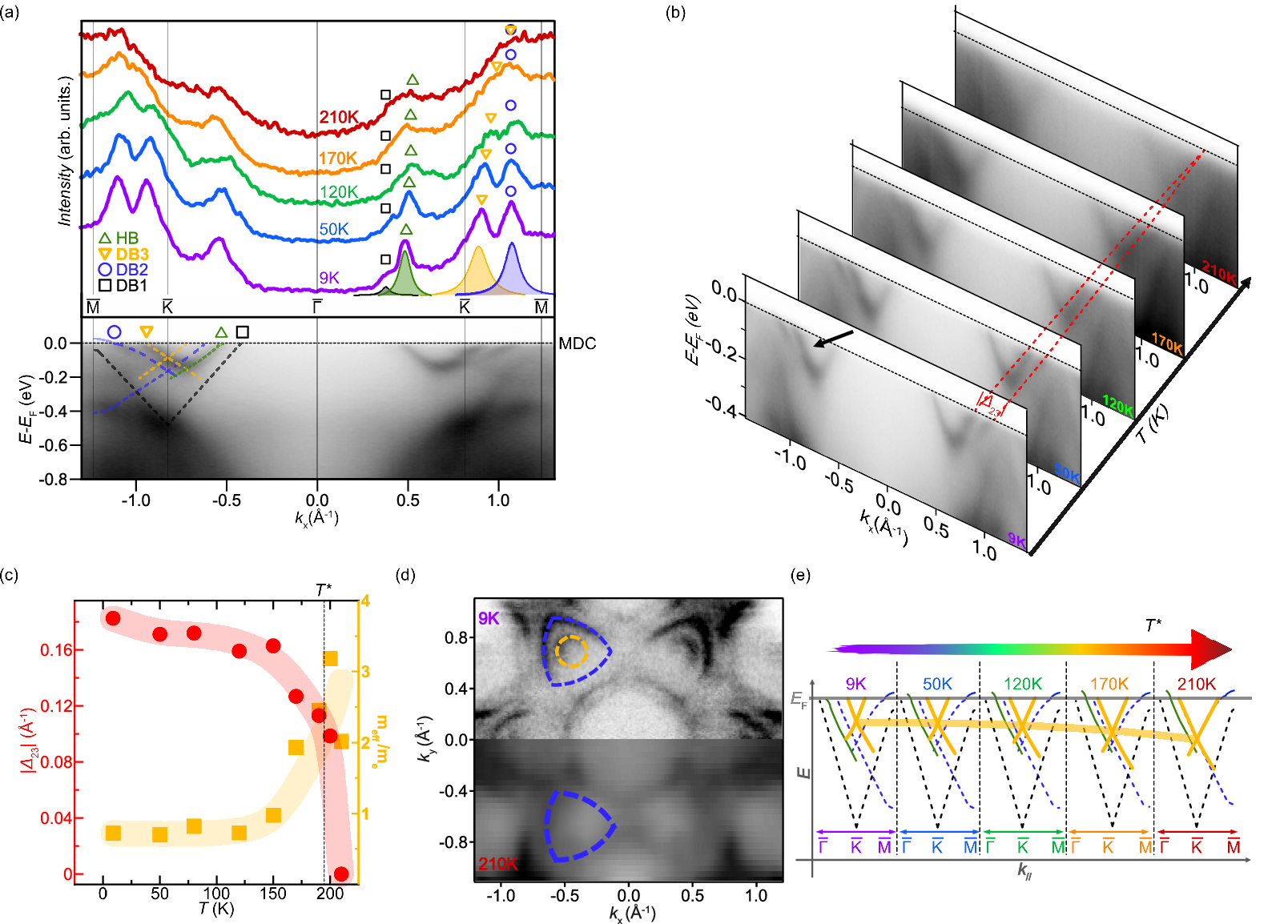}
    \centering
    \caption{\textbf{Temperature dependent ARPES data.}
    \textbf{(a)} The upper panel displays the temperature evolution of the momentum distribution curve (MDC) line profile taken at $E_F$ and the bottom panel is the spectral image of $\bar{M}-\bar{K}-\bar{\Gamma}-\bar{K}-\bar{M}$ cut at 9 K with LH polarization for a photon energy of 102 eV. The green triangle, yellow inverse triangle, blue circle, and black square symbols each mark the peak positions of the HB, and the Dirac band (DB)s 3, 2, and 1, respectively. \textbf{(b)} $\bar{M}-\bar{K}-\bar{\Gamma}-\bar{K}-\bar{M}$ cuts stacked along with respect to temperature. The red dotted lines are guides to the closing of band splitting ($\Delta_{23}$) as the temperature rises. The black arrow highlights the dispersion discontinuity. \textbf{(c)} The filled red circles represent the fitted $\Delta_{23}$ values while the filled yellow squares represent the effective mass of DB3. The vertical dotted lines of (c) mark \textit{T}*. \textbf{(d)} Temperature-dependent Fermi surfaces at 9 K and 210 K, presented in the $k_y$$>$0 and $k_y$$<$0 regions, respectively. \textbf{(e)} Schematic diagram of the band shift of DB3 and HB with respect to temperature. The black, blue, yellow, and green (dotted) lines in (a), (d), and (e) serve as guides to the eye for DBs 1, 2, 3, and HB, respectively.} 
    \label{fig3}
\end{figure*}

Moreover, we observed a transition displayed by the kink at $T^*$, which is absent in polycrystalline ScFe$_6$Ge$_6$ measurements~\cite{RFGref1}. $T^*$ is only shown on the in-plane result and not on the out-of-plane data (Fig.~\ref{fig1}(f)). This hints at the onset of a weak ferromagnetic component by the Fe spin canting away from the easy axis. Note that such spin-canting-like anomaly at low temperature is also observed in the parent compound FeGe~\cite{FG0,FG2MingNatphy,FG5-1pengnatcomm,FG5PengarX,FG9vacancypeng}, while an additional magnetic transition at $T^*$ appears in ScFe$_6$Ge$_6$ instead of a CDW phase as in FeGe.

The electronic band structure of ScFe${_6}$Ge${_6}$ kagome metal was obtained to gain a better understanding of the mechanism underlying this new transition at $T^*$. The ARPES spectra were collected at 9 K with a photon energy of 102 eV crossing the $\Gamma$ point (Fig.~\ref{fig2}(a), Fig. S2). All subsequent discussions are based on the two-dimensional projected Brillouin zone (BZ) because of the $k_z$ broadening effect. All three characteristic features of a kagome band can be identified, including the FB at $\approx -500$ meV, as observed in the spectral image of the high-symmetry cut (Fig. S2(b)).
 
The $\bar{K}-\bar{M}-\bar{K}$ cut (Fig.~\ref{fig2}(b)) was examined to elucidate the kagome band structure of ScFe${_6}$Ge${_6}$. Three sets of linearly dispersing Dirac bands are identified near the $E_F$: Three DCs (DC1, DC2, and DC3) and three VHSs (VHS2-t, VHS2-b, and VHS1) (Fig.~\ref{fig2}(b),(c)). Note that the hole band (HB), the green dotted line in Fig.~\ref{fig2}(b), originates from a circular hole pocket centered at the $\Gamma$ point with a large Fermi wave vector $k_F\approx0.5$\AA$^{-1}$ (Fig. S2(a),(c)).

The DCs can be better appreciated by examining the constant energy contour plot in Fig.~\ref{fig2}(d). The Fermi momenta of DC1 and DC3 form circular Fermi pockets (white and orange dotted lines), while the DC2 band forms a triangular pocket (blue dotted line), which is a common trait in kagome systems hosting VHSs near $E_F$ ~\cite{trian,FG2MingNatphy,Cs135,VHS1,VHS2,VHS3}. The circular and triangular contours gradually shrink toward the Dirac points at the $\bar{K}$ point, forming a conical dispersion, as shown in Fig.~\ref{fig2}(c) and (d). The crossing points of DC3, DC2, and DC1 correspond to energies of $-$85, $-$160, and $-$480 meV, respectively. 

The DC3 band, which lies closest to $E_F$, does not exhibit an apparent VHS, whereas the DC2 and DC1 bands each display distinct VHSs. Cuts parallel to $\bar{K}-\bar{M}-\bar{K}$ are stacked along the perpendicular $\bar{\Gamma}-\bar{M}-\bar{\Gamma}$ direction (Fig.~\ref{fig2}(d)-(f)) to confirm the saddle point-like dispersion (Fig.~\ref{fig2}(c)). The VHS2-t and VHS1 each exhibit band tops just above the $E_F$ and at $\approx -22$ meV, respectively (Fig.~\ref{fig2}(b)), while the VHS2-b displays a band bottom near $-$400 meV (Fig. S4(b)). Meanwhile, along the direction orthogonal to that shown in Fig.~\ref{fig2}(b), the band dispersions of VHS1 and VHS2-t exhibit positive curvature, with the band minima located at the $\bar{M}$ point, as indicated by the thick dashed lines in Fig.~\ref{fig2}(e),(f). The VHS2-b exhibits opposite concavity to VHS2-t and VHS1 (Fig.~\ref{fig2}(b), Fig. S4). 

The role of the electronic band structure in the magnetic phase transition at $T^*$ was next examined by taking temperature-dependent ARPES measurements along the $\bar{M}$--$\bar{K}$--$\bar{\Gamma}$--$\bar{K}$--$\bar{M}$ cut. Momentum distribution curve (MDC) fitting was performed to quantitatively assess the band evolution. The MDC curves were extracted at $E_F$ to resolve the Fermi wave vector $k_F$ of each band crossing the Fermi level (Fig.~\ref{fig3}(a)). The MDC line profiles were fitted using Lorentzian functions on a cubic background, and four peaks were identified below $T^*$. The HB (green triangle) and Dirac band DB1 (black square) peak positions lie in close proximity, with the DB1 intensity being substantially weaker than that of HB. The other two bands corresponding to DB2 (blue circle) and DB3 (yellow inverse triangle) are clearly visible near the $E_F$ in Fig.~\ref{fig3}(a). Note that with respect to the $\bar{K}$ point, each branch of DB2 and DB3 appears only on one side of the cut; DB2 and DB3 are visible only in the $\bar{K}-\bar{M}$ region. This accounts for why the combination of HB and DB3 appears, at a first glance, as a single band forming an electron pocket. However, a closer inspection of the spectral image at 9 K reveals a slight discontinuity in the band dispersion at the $\bar{K}$ point where the DBs cross, as highlighted by the black arrow in Fig.~\ref{fig3}(b). 

Since we have resolved the band structure, we look for any evolution of the electronic state as the temperature increases. The peak position difference ($\Delta_{23}$) between DB2 and DB3 decreases with increasing temperature, eventually closing above $T^*$, as shown in Fig.~\ref{fig3}(b) and (c).
It is noteworthy that the $\Delta_{23}$ gap reopens upon cooling back to the base temperature after a full heating cycle, excluding external origins of the evolution, such as surface contamination (Fig. S5). 

The nature of this gap closing of $\Delta_{23}$ is clearly resolved by comparing the Fermi surfaces in the ARPES data at 210 K and 9 K (Fig.~\ref{fig3}(d)). The circular Dirac cone DC3 located within the triangular DB2 is visible at 9 K, whereas at 210 K (above $T^*$), no spectral weight is observed inside DB2. Hence, band merging behavior exists, where only the $k_F$ of DB3 increases until it merges with DB2. The nature of DB2 and DB3 merging can also be elucidated by examining the effective mass of DB3, determined by the $k_F$ and the band slope. While the other kagome bands remain relatively rigid with increasing temperature, DB3 undergoes a pronounced change as the band is lowered, eventually merging with DB2, exhibiting a reduced slope indicative of an enhanced effective mass (Fig.~\ref{fig3}(c)). 
The schematic diagram in Fig.~\ref{fig3}(e) summarizes how DB3 (yellow line) merges with the rigid DB2. In particular, the schematics at 120 K and 200 K show a clear reduction in slope, whereas those at lower temperatures exhibit only minor slope changes, consistent with the behavior shown in Fig.~\ref{fig3}(c). 

Spin splitting associated with Stoner-type itinerant magnetism is a natural explanation for the merging of two bands at the magnetic transition temperature~\cite{FG2MingNatphy,mingyiMBE}. Nevertheless, the band geometries of DB2 and DB3 are fundamentally different, as evident from the Fermi surface (Fig.~\ref{fig3}(d)); the former exhibiting a triangular cone shape and the latter a circular cone shape (Fig.~\ref{fig2}(c)). Thus, a spin-split type scenario between DB2 and DB3 is unlikely. 

Rather, we speculate that the band shift of DB3 reflects an orbital-selective response arising from strong coupling between Fe spin order and electronic states derived from out-of-plane Fe $d$-orbitals. The changes in the spin configuration are expected to have a greater impact on electronic states with out-of-plane orbital geometry because the Fe moments are aligned out-of-plane in the A-type AFM phase. Namely, out-of-plane orbital bands are particularly sensitive to variations in out-of-plane magnetization and interlayer interactions~\cite{massivedirac3}. Indeed, the pronounced thermal evolution of DB3, compared to the other bands, across the magnetic transition $T^*$ suggests that spin reorientation selectively influences the Dirac fermions. Such orbital-selective evolution of Dirac bands was also reported in Fe$_3$Ge, another FeGe based compound~\cite{massivedirac3}.

In typical $d$-orbital kagome materials, the $d_{xy}$/$d_{x^2-y^2}$, $d_{xz}$/$d_{yz}$, and $d_{z^2}$ orbitals form three sets of kagome bands (also confirmed in orbital-resolved density functional theory (DFT) result in Fig. S3). Hence, it is natural to assign the three DBs (DB1, DB2, and DB3) to each set of orbitals, despite differences in details between the DFT results and the experimental data. Since DFT reveals the density of states near $E_F$ to mainly consist of Fe $d$-orbitals~\cite{RFGref3}, DB3 should likewise originate from one set of the kagome Fe-3$d$ orbitals.

\begin{figure*}
    \includegraphics[width=16cm]{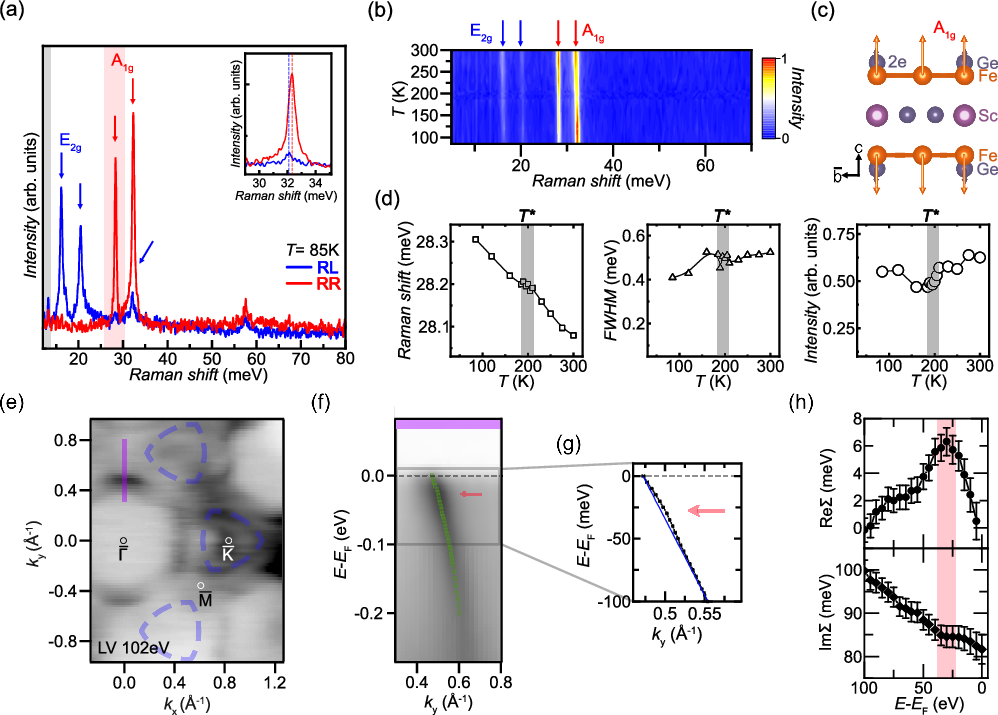}
    \centering
    \caption{\textbf{Raman data of ScFe$_6$Ge$_6$ and the corresponding kink of the HB in ARPES result.}
    \textbf{(a)} Raman spectra in the RL and RR polarization at 85 K. The inset zooms into 32 meV to highlight the energy difference between the $A_{1g}$ and $E_{2g}$ phonon modes. \textbf{(b)} Color contour plot of the temperature dependence in unpolarized symmetry. The blue and red arrows in (a) and (b) mark the $E_{2g}$ and $A_{1g}$ phonons, respectively. \textbf{(c)} Vibrational mode of the $A_{1g}$ phonon mode at 28 meV (A-mode). \textbf{(d)} Temperature dependence of the energy, FWHM, and intensity of the A-mode. \textbf{(e)} Fermi surface of ScFe$_6$Ge$_6$ at 20 K measured in the LV geometry. The blue broken triangles mark the DB2 band. \textbf{(f)} The ARPES intensity plot corresponding to the purple line in (e). The band dispersion fitted from the MDCs is overlaid as green squares. \textbf{(g)} Zoomed-in result of the MDC fitting. The blue linear line represents the bare-band to emphasize the kink feature. The red arrows of (f) and (g) point to the kink. \textbf{(h)} The upper panel shows the real part of the electron self-energy ($\Sigma$) obtained from the MDC peak position. The lower panel displays the imaginary part of $\Sigma$ derived from the width of MDC fitting. The kink position corresponding to the A-mode is marked with the red vertical line in (a) and (h).} 
    \label{fig4}
\end{figure*}

In particular, to identify which orbital contributes to DB3, we compare ARPES cuts along the $\bar{\Gamma}-\bar{K}-\bar{M}$ direction at 9 K under LH and LV polarizations (Fig. S6). In our experimental setup, the plane of incidence is aligned parallel to a sample mirror plane, such that LH (LV) corresponds to p- (s-) polarization, predominantly probing orbitals with even (odd) symmetry with respect to the mirror plane~\cite{ARPESrevdama,ARPESrev,polgeome}. Indeed, DB3 is observed only under LH polarization, suggesting that it is dominated by orbitals with even symmetry and strong out-of-plane character such as the $d_{z^2}$ orbital. This assignment is consistent with our argument that the out-of-plane orbital component is strongly coupled to the $T^*$ transition. Considering that the $T^*$ transition corresponds to spin canting away from the c-axis, the resulting influence is most likely to be captured in the out-of-plane derived DB3.

The magnitude of the effective mass renormalization of DB3 across $T^*$ provides additional insight into the origin of the temperature-dependent evolution. The light effective mass (0.7$m_e$) at 9 K increases by more than fourfold (3.2$m_e$) at 200 K. Such strong carrier renormalization indicates enhanced scattering arising from mechanisms such as change in spin–orbit coupling (SOC), in orbital hybridization or in inter-layer coupling associated with the possible spin reorientation at $T^*$. We remark that from the discontinuity marked by the black arrow in Fig.~\ref{fig3}(b), the opening/closing of a SOC-induced gap can be a possible origin of the mass renormalization accompanying the magnetic transition ~\cite{massivedirac1,massivedirac2,massivedirac3}. Nevertheless, within the resolution of our data, such a gapped feature cannot be conclusively identified and calls for further studies via high-resolution ARPES.

Having addressed the electronic structure, we will next examine the crystallographic properties. In FeGe, the lattice degrees of freedom have been found to play a key role in determining its rich charge and spin ordered phases~\cite{FG12hundmetal,FG10Raman}. The phonon modes in ScFe${_6}$Ge${_6}$ were assigned by Raman spectroscopy within the $ab$ plane of ScFe${_6}$Ge${_6}$ using circularly polarized light (RL and RR channels) at 85 K. All Raman-active phonon modes obtained with the factor group analysis for the space group P6/mmm (2$A_{1g}$+3$E_{2g}$) were observed and assigned. The RR and RL scattering geometries access the $A_{1g}$ and $E_{2g}$ channel, respectively~\cite{FG10Raman}. Hence, the three peaks observed at 16.2, 20.5, and 32.1 meV (marked by the blue arrows in RL polarization) correspond to the three $E_{2g}$ phonons, and the peaks at 28.3 and 32.3 meV (marked by the two red arrows) are assigned to $A_{1g}$ symmetry via DFT calculations (Table S1). The mode near 32 meV consists of two peaks separated by 0.2 meV (inset of Fig.~\ref{fig4}(a)), confirming that these are two distinct phonons rather than signal leakage caused by poor polarization control. We note that the weak peak near 57 meV observed in LR and RR configurations can be attributed to $E_{1g}$ phonon leakage resulting from a slight misalignment from the $c$-axis.

Next, temperature-dependent Raman scattering measurements were taken to elucidate the role of lattice degrees of freedom in the transition at $T^*$ (Fig. S7). The color contour plot in Fig.~\ref{fig4}(b) illustrates the temperature evolution of the dominant phonon modes measured in an unpolarized geometry. No significant changes are observed in the overall phonon modes, such as the emergence or disappearance of new modes, indicating the absence of a structural phase transition or lattice distortion. This result is in stark contrast with the temperature-dependent Raman results for FeGe, which showed a complex sequence of structural distortions as the temperature decreased~\cite{FG10Raman}.

Despite the overall absence of dramatic temperature dependent behavior across all phonon modes, the $A_{1g}$ phonon at 28 meV (A-mode), arising from the in-phase out-of-plane vibrations of the Fe kagome plane and Ge atoms on the 2$e$ Wyckoff positions (Fig.~\ref{fig4}(c)), exhibits an anomaly near $T^*$ (Fig.~\ref{fig4}(b) and (d)). A discontinuous kink-feature is observed for the A-mode frequency at $T^*$ (Fig.~\ref{fig4}(d)), marking a deviation from a simple anharmonic behavior. In addition, the full width at half maximum (FWHM) broadens with increasing temperature, before exhibiting constant behavior above $T^*$. These anomalies are accompanied by an intermediate intensity dip around $T^*$, pointing to renormalization of the phonon.

Such anomalies of the A-mode phonon point to a magnetoelastic effect, namely spin–phonon coupling. In contrast to other phonon modes, such as the in-plane $E_{2g}$ modes, the out-of-plane A-mode serves as a natural local probe of the interlayer AFM spin configurations~\cite{magnetophonon}, which are governed by the modulation of the interlayer exchange interaction and single-ion anisotropy between the ferromagnetic kagome layers. Moreover, the in-phase vibrations of the A-mode are expected to be more sensitive to magnetic ordering than their out-of-phase counterparts, because they coherently modulate the Fe–Ge–Fe exchange interaction pathways. Indeed, recent first-principles calculations showed that certain phonon modes are significantly affected by the magnetic order, whereas conversely, the magnetic interactions are also dramatically influenced by the structural environment in Fe-based kagome systems, supporting strong magnetoelastic coupling effects in FeGe~\cite{magnetophonon_Mazin}. Thus the resulting anomalies in frequency, linewidth, and intensity across $T^*$ provide direct evidence of spin–phonon coupling associated with the reconfiguration of the magnetic ground state. 
In this context, $T^*$ likely signifies spin canting behavior within the A-type AFM regime. However, the coexistence of electron-phonon and spin-phonon coupling obstructs the exact origin of the subtle anomalous Raman behavior at the transition $T^*$. More theoretical studies are needed to resolve this open issue.

Phonon anomalies at $T^*$ occur concurrently with modifications in the electronic band structure, as shown in the ARPES results above. The bands in close proximity to $E_F$ were examined to determine the strength of electron-phonon coupling. The presence of kinks in band dispersions indicates electrons coupled to bosonic modes such as phonons~\cite{ARPESrev,ARPESrevdama}.
Indeed, we observe a kink at $\approx$28 meV in the real and imaginary parts of the electron self-energy (Fig.~\ref{fig4}(e)--(h)). The green square kink positions in Fig.~\ref{fig4}(f) were obtained with MDC fitting with a Lorentzian peak. This shows the coupling of electrons to a bosonic mode at 28 meV below the $E_F$. Surprisingly, the energy of the A-mode in the Raman results matches this energy. A coupling constant of $\lambda \approx 0.3$, comparable to that of FeGe, was obtained by extracting the ratio of the slope from the real part of the self-energy~\cite{FG2MingNatphy,ARPESrevdama}. This indicates moderate coupling of the $A_{1g}$ phonon to the electronic sector. Moreover, as shown in Fig.~\ref{fig3}(a) and (e), the involved HB also exhibits an upward energy shift with increasing temperature, accompanied by an increase in $k_F$. The similar temperature sensitivity of the HB and the A-mode further supports the presence of electron–phonon coupling in ScFe$_6$Ge$_6$.

\begin{center}
\noindent\textbf{III. DISCUSSION and CONCLUSION}
\end{center}
Finally, we conclude with a one-to-one comparison between ScFe$_6$Ge$_6$ and its parent compound FeGe across the relevant degrees of freedom. The absence of CDW indicates that the charge degrees of freedom in ScFe$_6$Ge$_6$ play a weaker role than in FeGe. Temperature-dependent measurements of the evolution of resistivity, magnetism, electronic band structure, and Raman spectrum revealed pronounced anomalies at the CDW ordering temperature of FeGe~\cite{FG1Nat,FG2MingNatphy,FG10Raman}. On the other hand, no CDW transition is observed in ScFe${_6}$Ge${_6}$. In contrast to FeGe having CDW gaps in the VHS bands~\cite{FG2MingNatphy,FG4SciadMing}, ScFe$_6$Ge$_6$ does not show any sign of gap opening. Interestingly, although the VHSs are located close to the $E_F$ for both FeGe and ScFe$_6$Ge$_6$, no electronic instability related phenomena are observed in ScFe$_6$Ge$_6$, indicating a diminished role of VHSs in the formation of CDW in FeGe-related compounds. The absence of CDW is also supported by the absence of phonon instability in ScFe$_6$Ge$_6$ (Fig. S8) contrary to FeGe~\cite{FG10Raman}.

The role of lattice degrees of freedom in ScFe$_6$Ge$_6$ is also reduced compared to FeGe. The electron-phonon coupling ($\lambda$$\approx$0.3) of ScFe$_6$Ge$_6$ obtained from kink analysis is smaller than that of FeGe ($\lambda$$\approx$0.5)~\cite{FG2MingNatphy}. In addition, the absence of dramatic change in the fitted phonon parameters, contrary to FeGe~\cite{FG10Raman}, also points to a weaker spin-phonon coupling. Taken together, these results suggest that the lattice degrees of freedom are more weakly coupled to the other degrees of freedom in ScFe$_6$Ge$_6$.

On the other hand, the out-of-plane crystal environment still appears to play an important role on its own. A qualitative argument can be made based on the crystal structure of ScFe$_6$Ge$_6$. Since the distance between neighboring Fe kagome layers is longer for ScFe$_6$Ge$_6$ compared to FeGe, the Ge atoms are being further separated by the Fe plane. Such separation would disrupt the dimerization (or interaction) of the out-of-plane Ge atoms, leading to the absence of CDW ordering. Considering the fragility of the CDW in the presence of Ge vacancies~\cite{FG4SciadMing,FG5PengarX,FG9vacancypeng,FG12hundmetal}, our findings in ScFe$_6$Ge$_6$ are in line with the consensus that the CDW is strongly correlated with Ge dimerization.

Lastly, the orbital degree of freedom, which apparently plays a small role in FeGe, becomes a prominent factor in ScFe$_6$Ge$_6$~\cite{FG4SciadMing,FG10Raman}. Orbital-selective band evolution correlated with magnetism, as observed in ScFe$_6$Ge$_6$, has not been reported in FeGe. These findings indicate that in ScFe$_6$Ge$_6$, with spin degrees of freedom taking the central role, greater weight is placed on orbital degrees of freedom than on charge and lattice degrees of freedom.
Adding to this argument, we note that the orbital-selective behavior of the band evolution, especially in the out-of-plane 3$d$-orbital, and the correlation with the $A_{1g}$ phonon mode suggests that modulation/interaction along the c-axis strongly influences the properties of ScFe$_6$Ge$_6$. The spin-canting or spin density wave~\cite{FG5PengarX,FG5-1pengnatcomm,FG9vacancypeng} and CDW ordering in FeGe are likewise highly sensitive to the c-axis interactions and geometry. 
The shared importance of c-axis fluctuations in both ScFe$_6$Ge$_6$ and FeGe points to a common mechanism, rooted in out-of-plane correlations, that governs the behavior of FeGe-related compounds.

In summary, we have probed the lattice, spin, charge, and orbital degrees of freedom using complementary characterization and spectroscopy techniques. The magnetic phase transitions present in single-crystal ScFe${_6}$Ge${_6}$ were confirmed, and the corresponding temperature-dependent anomalies of the phonon modes and electronic kagome band structure were studied. Although no structural symmetry breaking or long-range charge order is observed, a secondary magnetic transition at $T^*$ within the A-type AFM state coincided with an anomaly in the $A_{1g}$ phonon suggesting a magnetoelastic response, indicative of a subtle canting of Fe spins~\cite{magnetophonon_Mazin}. Concurrently, ARPES measurement showed temperature-dependent orbital-selective renormalization of a Dirac band, along with a kink feature at the $A_{1g}$ phonon energy. Such results show that ScFe$_6$Ge$_6$, and FeGe-compounds in general, make up a complex system where multiple degrees of freedom are intricately correlated via the out-of-plane interaction and modulation. More broadly, our study of a FeGe-based magnetic system provides a stepping stone toward elucidating the mechanisms that enable or suppress CDW order in the presence of magnetic order, while emphasizing the enhanced role of orbital degrees of freedom in stabilizing such spin-correlated charge order.

\begin{center}
\noindent\textbf{IV. METHODS}
\end{center}

\noindent\textbf{Sample preparation and characterization of physical properties}

ScFe$_6$Ge$_6$ single crystals were grown using the Sn-flux method. A mixture of Sc ingot (99.95\%), Fe powder (99.99\%), Ge block (99.99\%), and Sn granules (99.99\%) in a molar ratio of 1:3:6:20 was sealed in an alumina crucible under vacuum within a quartz ampoule. The sealed ampoule was heated to 1100\,${^{\circ}}C$ over 20\,$h$, held for 24\,$h$, and then slowly cooled to 700\,${^{\circ}}C$ at a rate of 2\,${^{\circ}}C/h$. Excess Sn flux was removed at this temperature by high-speed centrifugation, after which the ampoule was cooled to room temperature in air. The residual surface flux was eliminated by briefly dipping the crystals in dilute hydrochloric acid.

SXRD confirmed that ScFe$_6$Ge$_6$ crystals adopt a hexagonal structure (space group P6/mmm, No. 191). Structural refinement gave lattice parameters of $a = b =$ 5.10\,{\AA}, $c =$ 8.12\,{\AA}, and $\alpha$ = $\beta$ = 90${^{\circ}}$, $\gamma$ = 120${^{\circ}}$, The sharp reciprocal lattice patterns exhibited the high crystallinity of the obtained crystals (Fig. S1).

The magnetic susceptibility data were measured using a superconducting quantum interference device magnetometer (MPMS3, Quantum Design). ZFC and FC temperature dependent dc magnetic susceptibility curves were obtained at a field of 100 Oe with the sample $ab$-plane aligned parallel or perpendicular to the field. 

\noindent\textbf{Spectroscopy measurements}

High-resolution $\mu$-angle-resolved photoemission spectroscopy (ARPES) measurements were taken at the BL03U beamline of Shanghai Synchrotron Radiation Facility (SSRF)~\cite{yang2021high}. All ARPES data were acquired with a Scienta-Omicron DA30 analyzer using 102~eV photons. The total energy resolution was better than 20~meV, and the angular resolution was set to 0.2$^\circ$. High-quality single crystals were cleaved \textit{in situ} under ultrahigh vacuum, and the base pressure in the main chamber was maintained below $1\times10^{-10}$ torr. The LV-polarization data were denoised using a deep-learning-based convolutional neural network~\cite{CNN}.

Raman scattering experiments were carried out using a $\lambda = 515$ nm laser (Cobolt Fandango) with a laser power of less than 0.5 mW at the sample position and a spot diameter of about 2 $\mu$m to reduce laser heating. The temperature-dependent measurements were performed in an open-flow microscope-type cryostat (Oxford MicroStat HR). Scattered light was dispersed through a TriVista 777 (Princeton Instruments) spectrometer, while the data were recorded using a liquid-nitrogen-cooled charge-coupled device detector (PyLoN eXcelon).

\noindent\textbf{Band calculation}
The electronic structure of ScFe$_6$Ge$_6$ is calculated using the DFT method \cite{1}, with the EDMFTF code developed by Haule $et$ $al.$ \cite{2}, which builds on the full-potential linear augmented plane wave method implemented in WIEN2k \cite{3}. A fully charge-self-consistent procedure is performed in DFT calculation. The calculations are performed at T= 290\,K. The experimental crystal structure (space group P6/mmm, No. 191) of ScFe$_6$Ge$_6$ with lattice constants $a = b =$ 5.053\,{\AA} and c = 8.138\,{\AA} \cite{RFGref1,RFGref2,RFGref3} is used in the calculations. The orbital contributions were calculated within the framework of density functional theory (DFT) using the projector augmented wave method \cite{10}, as implemented in the Vienna Ab initio Simulation Package (VASP) \cite{11}. The Perdew-Burke-Ernzerhof (PBE) generalized gradient approximation \cite{4} was used for the exchange-correlation functional. A cutoff energy of 450 eV was applied, and Brillouin zone integration was performed using a Monkhorst-Pack k-point mesh of 12$\times$12$\times$6. Post-processing was carried out using the VASPKIT program \cite{12}. Phonon dispersion and Raman-active modes were then calculated with PHONOPY \cite{13} using a 2$\times$2$\times$1 supercell, a 6$\times$6$\times$6 $k$-point mesh, and a 450\,eV cutoff energy.

\section*{Acknowledgments}
The work by J.H.L. and C.K. was supported by Global Research Development Center (GRDC) Cooperative Hub Program through the National Research Foundation of Korea (NRF) funded by the Ministry of Science and ICT(MSIT) (Grant No. RS-2023-00258359) and the NRF grant funded by MSIT (Grant No. NRF-2022R1A3B1077234). This work was also supported by the Institute of Applied Physics, Seoul National University. D.W. acknowledges support from the ITRC Program through the IITP, and from the Global Research Development Center (GRDC) Cooperative Hub Program through the NRF, funded by the (MSIT) (Grant Nos. RS-2024-00437191 and RS-2023-00258359). The authors also thank the Analytical Instrumentation Center (\#SPST-AIC10112914) and the Double First-Class Initiative Fund of ShanghaiTech University.


\section*{Competing interests}
The authors declare no competing interests.

\section*{Data availability}
The data that support the findings of this study are available from the corresponding authors upon reasonable request.

\bibliography{agsi-bibliography}

\end{document}